\documentclass[conference]{IEEEtran}
\IEEEoverridecommandlockouts

\usepackage{cite}
\usepackage{url}

\usepackage{nameref,hyperref}
\usepackage{amsmath,amssymb,amsfonts}
\usepackage{algorithmic}
\usepackage{graphicx}
\usepackage{textcomp}
\usepackage[table,xcdraw]{xcolor}
\def\BibTeX{{\rm B\kern-.05em{\sc i\kern-.025em b}\kern-.08em
		T\kern-.1667em\lower.7ex\hbox{E}\kern-.125emX}}

\usepackage{multirow}

\usepackage{listings}
\usepackage{color}

\usepackage[table,xcdraw]{xcolor}
\usepackage[normalem]{ulem}
\useunder{\uline}{\ul}{}

\definecolor{dkgreen}{rgb}{0,0.6,0}
\definecolor{gray}{rgb}{0.5,0.5,0.5}
\definecolor{mauve}{rgb}{0.58,0,0.82}

\begin{document}

\title{Machine Learning Model Development from a Software Engineering Perspective: \\A Systematic Literature Review}

\author{
	\IEEEauthorblockN{Giuliano Lorenzoni}
	\IEEEauthorblockA{\textit{David R. Cheriton School of Computer Science} \\
	\textit{University of Waterloo (UW)}\\
	Waterloo, Canada \\
	glorenzo@uwaterloo.ca}
	\and
	\IEEEauthorblockN{Paulo Alencar}
	\IEEEauthorblockA{\textit{David R. Cheriton School of Computer Science} \\
		\textit{University of Waterloo (UW)}\\
		Waterloo, Canada \\
		palencar@uwaterloo.ca}
	\and 
    \IEEEauthorblockN{\hspace{2cm}Nathalia Nascimento}
	\IEEEauthorblockA{\hspace{2cm}\textit{David R. Cheriton School of Computer Science} \\
	\hspace{2cm} \textit{University of Waterloo (UW)}\\
	\hspace{2cm}	Waterloo, Canada \\
	\hspace{2cm} nmoraesdonascimento@uwaterloo.ca}
	\and
	\IEEEauthorblockN{\hspace{2cm}Donald Cowan}
	\IEEEauthorblockA{\hspace{2cm}\textit{David R. Cheriton School of Computer Science} \\
	\hspace{2cm} \textit{University of Waterloo (UW)}\\
	\hspace{2cm} Waterloo, Canada \\
	\hspace{2cm} dcowan@uwaterloo.ca}
}

\maketitle

\begin{abstract}
Data scientists often develop machine learning models to solve a variety of problems in the industry and academy but not without facing several challenges in terms of Model Development. The problems regarding Machine Learning Development involves the fact that such professionals do not realize that they usually perform ad-hoc practices that could be improved by the adoption of activities presented in the Software Engineering Development Lifecycle. Of course, since machine learning systems are different from traditional Software systems, some differences in their respective development processes are to be expected. In this context, this paper is an effort to investigate the challenges and practices that emerge during the development of ML models from the software engineering perspective by focusing on understanding how software developers could benefit from applying or adapting the traditional software engineering process to the Machine Learning workflow. 


\end{abstract}


\section{Introduction}\label{sec:introduction}

In Software Engineering (SE), researchers and practitioners have spent decades on developing tools and methodologies to create, manage and assemble complex software modules. Software engineering refers to the comprehensive study of engineering for the design, development and maintenance of software with the main purpose of developing methods and procedures for software development, for example to scale it up for large systems and guarantee high-quality software with low cost software production. Indeed, SE has progressed beyond expectations to produce significant advances in its methods and processes. Therefore, in the light of these advances, it becomes important to understand how software developers could adapt their existing processes to incorporate or adapt the existing SE process to their Machine Learning (ML) workflows.

In this context, we investigate the challenges and practices that emerge during the development of ML models from the software engineering perspective.  By focusing our analysis on the well-known stages presented in the Software Engineering development process, we focus on investigating how software developers could benefit from applying or adapting these processes to the ML workflow since one might argue that data scientists would benefit from adopting classical software engineering disciplines (e.g. systems design, quality assurance, and verification) to build their models properly.

Consequently, this article is organized in 7 sections: (i) Section 2 regards with the related work; (ii) Section 3 covers the research method by describing: the research questions, the document source used to select the articles, the  search string/strategy and the inclusion/exclusion criteria; (iv) Section 4 covers the result analysis including demographics and the answers for each research question; (v) Sections 5 and 6 demonstrate the challenges for future research and the threats to the validity of the results found in this article. Finally (vi) the conclusions and insights from this research can be found in section 7.

\section{Related Work}

Amershi et al. \cite{8804457} conducted a comprehensive study involving AI professionals at Microsoft. By investigating scientists, researchers, managers, programmers, and other professionals in their respective daily activities, the authors have identified the three major challenges in building large-scale AI applications: data management, reuse, and modularity. In order to address the challenges for ML, the paper analyzes how Microsoft software teams adapt their existing agile processes to incorporate the complexity from machine learning to then build artificial intelligence-based applications such as text, voice, and video translators or the interactive speaking agents built on speech and language recognition. The authors interviewed company employees to find out how they tackle challenges from a development standpoint - especially in machine learning modeling. The paper contributes to the field by providing insights related to the adoption of software engineering processes to machine learn modeling, such as: (i) machine learning modeling dependency on data and the consequent importance of all data related stages in a machine learning modeling workflow; and (ii) the need for teams with skills for software engineering and machine learning modeling; (iii) the difficulty of modularization and reuse of machine learning models (which is different from regular software applications) since modules in machine learning have greater impact on each other. Consequently, by acknowledging that ML applications are structurally different from applications in other domains, the authors were able to build a comprehensive and up-to-date workflow (which was adopted in our present work as a common reference to understand the ML modeling process) with typical stages that address the ML model development.

While Amershi et al. \cite{8804457} identifies the main challenges regarding building large-scale AI applications and used it for building an up-to-date ML workflow, another comprehensive study by Correia et al. \cite{Correia} conducted interviews with data scientists from five different Brazilian companies in order to identify which were the most challenging stages of the ML workflow proposed by Amershi et al. \cite{8804457}. The authors found that data scientists have pointed to Data Processing and Feature Engineering as the most challenging stages in the ML workflow, even while they also mentioned important issues regarding Model Training, Model Evaluation, and Model Deployment. These results indicate the lack of a well-engineered process in the ML model development practice.


A study conducted by Zhang et al. \cite{zhang2019empirical} investigated 138 research papers in search of methods for testing and debugging the ML code. Their findings show that only a few contributions focus on testing interpretability, privacy, or efficiency. Zhang et al. \cite{zhang2019empirical} focus on exclusively analyzing the stage of Model Evaluation, as opposed to targeting all ML stages.

In contrast to the partial information in works such as Amershi et al. \cite{8804457}, authors such as Hesenius et al. \cite{hesenius2019towards} argue that although there are challenges faced by software engineers when developing data-driven applications, the data dependency of ML/AI applications does not constitute an issue to the adoption of a common integrated Software Engineering (SE) process, upon which the project's overall success would depend. As a consequence, by defining a set of roles (Software Engineer, Data Scientist, Data Domain Expert, and Domain Expert), stages, and responsibilities to structure the necessary work, decisions and documents, the authors provided a structured engineering process that suits all data-driven applications, ultimately filling the gap found in the literature. It is worth mentioning that although the article presents a general framework in contrast to specific solutions aiming to solve particular and partial issues related to ML/AI modeling, the adoption of the proposed model does not disregard specializations of the aforementioned process, including individual steps and specifically tailored tools for certain data-driven applications.

In a similar fashion, by presenting methods for measuring the best practices degree of adoption when investigating the relationship between different groups of practices and assessing/predicting their effects by performing regression models, Serban's \cite{serban2020adoption} article reaches conclusions that are in line with Hesenius et al.\cite{hesenius2019towards} in a way that there is a set of best practices which is applicable to any ML application development, regardless of the type of data under consideration. Additionally, the author has contributed to the evolution of such practices by presenting a methodology in which each practice is related to its effects and adoption rate.

Washizaki et al. \cite{washizaki2019studying} presented a related work which clearly complements both works of Hesenius et al.  \cite{hesenius2019towards} and Serban et al.\cite{serban2020adoption}. The paper addresses the classification of software engineering design patterns by conducting a systematic study which collects, classifies, and analyzes SE architecture and the design of ``bad" patterns for ML systems which links traditional software systems and ML systems architecture and design. One interesting result presented by the article is the understanding that SE patterns for ML systems are divided between two processes: The Machine Learning pipeline and SE development.

Between these two very distinct approaches, one regarding customized ML workflows related to the specifics of each ML based system and the other oriented towards a broadly general workflow (which would be suitable to any ML application regardless of its specifics),  there are several studies with different approaches for the relation/application of software engineering development process to ML modeling. Some of them deal with specifics like the adoption of software engineering best practices related to the development of application programming interfaces (Reimann, Kniesel-Wünsch \cite{reimann2020achieving}), while others address the accountability gap in ML/AI by proposing a framework based on software development best practices (Hutchinson et al.\cite{hutchinson2020towards}). 

Further, Nascimento et al. \cite{de2019understanding} pointed out that the differences between Traditional systems and Machine Learning systems can be identified by observing the differences between their respective software development activities. In fact, the authors identified that SE activities are more challenging for ML systems which follow specific four-stage software development process, namely: understanding the problem, handling data, building models and monitoring those models.


Other works focus on well-known stages of SE lifecycle or the ML Workflow by identifying and addressing many different types of gaps, such as in the works presented in \cite{IEEE7,IEEE17,ACM179,ACM173,ACM196,ACM43,ACM30}.

Finally, 
we have also found papers which demonstrate the difficulty with reconciling Software Engineering Development with Machine Learning Modeling due to the fundamental differences of such processes as the work of Kim \cite{IEEE14}. 

\section{Research Method} \label{sec:research}

To systematize the aforementioned knowledge, we conducted a systematic literature review based on the guidelines described in \cite{keele2007guidelines}. Our review protocol includes:  (i) the selection of the digital libraries; (ii) the definition and validation of the search string; (iii) definition of the inclusion/exclusion criteria; and (iv) the application of snowballing. 

Following this protocol, two researchers performed a parallel search in order to identify studies that address the research questions. Before including the papers to the final result collection, they evaluated and interpreted the papers by discussing their relevance and the possible answers (findings) for the research questions. 





\subsection{Research Questions}\label{sec:questions}

The main question regards how the adoption of Software Engineering Development process and practices could address the issues from Machine Learning Modeling. In order to answer that we designed five research questions:
\begin{itemize}
	\item RQ1: What are the phases addressed in terms of the machine learning model development?
	\item RQ2: What are the techniques applied in each of phases of the machine learning model development phases?
	\item RQ3: What are the pros and cons of each machine learning model development technique?
	\item RQ4: What are the gaps in terms of the model development lifecycle?
	\item RQ5: What are the trends regarding the techniques applied in the machine learning model development lifecycle?
\end{itemize}

RQ1 main objective is to establish common ground in terms of what phases are addressed in the ML model development. By identifying these stages, we can address which techniques are applied in each stage (RQ2) as well as identify different model development techniques and their advantages/disadvantages (RQ3) and eventual gaps in the model development life-cycle of ML models (RQ4). The identification of the latest trends regarding techniques applied in the machine learning model development life-cycle (RQ5) can also provide important insights related to the main question. 

\subsection{Document Sources}\label{sec:source}

The procedure to select the sources used for our systematic literature review starts with a choice of a well-defined document sources in the field: in this case we selected IEEExplore and ACM Digital Library. After defining the document sources, we refined the results by identifying the main venues for publishing research in ML and/or SE, mainly based on the H-index.
However, as it is an emerging topic, we also included workshops that are associated to conferences that are important in the respective communities, such as ICSE workshops.  

\subsection{Search Strategy}\label{sec:strategy}

The Search Strategy consists in applying a selected search string,
filtering out papers based on the Inclusion/Exclusion Criteria.
The final results inclusion comes from the adoption of the Snowballing method. Given the novelty of the subject we have adopted a more comprehensive/general search string in order to get different combinations from the selected key words:  

\begin{quote}
    \itshape{ Title:(machine AND learning AND software AND engineering) OR Abstract:(machine AND learning AND software AND engineering)}\footnote{IEEExplore uses different reserved words for document title and abstract.}
\end{quote}

After the automatic search with this search string, we collected a total of 863 papers\footnote{The full list of the papers  is available at \href{https://drive.google.com/drive/folders/1nxivBZ8Kyh2zS2NmuoBnUiQ8znGB1v8O?usp=sharing}{Drive - List of Papers}}: 539 from IEEE and 381 from ACM (having a overlapping of 57 papers). First, we reduced the collection of articles by filtering them according to the venue title. 
Then, we excluded items that satisfy any of the exclusion criteria, such as short papers. Finally, we read the abstract of each paper, evaluating whether it satisfies the inclusion criteria.   As we summarized in Table \ref{table:searchresult}, of the 863 papers selected with the automatic string match, only 53 papers directly contribute to our research questions. Of the 53 papers, we selected the 23 most relevant. We observed that the greater part of the papers was filtered out because they describe solutions of ML to SE, as in the article of Nascimento et al. \cite{nascimento2018toward}. 


\begin{table}[!htb]
\caption{Summary of the Search Results.}
\label{table:searchresult}
\centering
\scalebox{0.95}{
\begin{tabular}{c|c|c|c|c|c|}
\cline{2-6}
                            & \multirow{2}{*}{\begin{tabular}[c]{@{}c@{}}Automatic Search\\ (2010-2020)\end{tabular}} & \multicolumn{4}{c|}{\begin{tabular}[c]{@{}c@{}}Selected\\ Papers\end{tabular}} \\ \cline{3-6} 
                            &                                                                                         & \textless{}2018          & 2018-2019          & 2020          & Total          \\ \hline
\multicolumn{1}{|c|}{IEEE}  & 539                                                                                     & 0                        & 10                 & 8             & 18             \\ \hline
\multicolumn{1}{|c|}{ACM}   & 381                                                                                     & 1                        & 2                  & 2            & 5             \\ \hline
\multicolumn{1}{|c|}{Total} & \begin{tabular}[c]{@{}c@{}}863 \\ (overlapping of 57)\end{tabular}                      & 1                        & 12                 & 10            & 23             \\ \hline
\end{tabular}
}
\end{table}

\subsection{Inclusion and Exclusion Criteria}

We consider papers that do not satisfy any of the exclusion criteria and satisfy at least two inclusion criteria. Thus, we excluded items that:
\begin{itemize}
	\item Papers written in languages other than English.
	\item Tutorials, short papers, editorials because they do not contain sufficient data for our study.
	\item Items related to Machine learning for software engineering.
\end{itemize}

We include studies that:  
\begin{itemize}
	\item Were published from January 2010 to June 2020. 
	\item RQ1 and RQ2: Have abstracts or document titles which mention/discuss the adoption of Practices/Processes/Practices/Workflow/Framework from Software Engineering for Machine Learning Modeling/Systems/Applications.
	\item RQ3 and RQ4: Matched the focus of the study (understanding how software teams could benefit from applying/adapting the traditional software engineering process to the Machine Learning workflow).
\end{itemize}

Finally, we have adopted the Snowballing method for refining our results based on the citations and references in the most relevant articles. For example, we have included some articles that were cited by Amershi et al \cite{8804457} and also some articles that have Amershi et al. \cite{8804457} as one of their references. Because of the novelty of the research topic, we considered grey literature that has already been cited. This step resulted on the addition of 10 papers.

\section{Result Analysis}

\subsection{RQ1: What are the phases addressed in terms of the machine learning model development?} \label{sec:RQ1}
The main purpose is to adapt or integrate Machine Learning framework into Software Development processes' stages, namely: requirements, design, implementation, testing, deployment, and maintenance. In this context, although there are some papers such as Nascimento et al. \cite{de2019understanding} that have developed Machine Learning Model workflows, the work of Amershi et al. \cite{8804457} presented the most comprehensive and accepted Machine Learning workflow which was mentioned and used in other articles (such as Correia et al. \cite{Correia}). The stages addressed in terms of Machine Learning Model Development were: 

\begin{itemize}
	\item A Model requirements stage which is related to the agreement between stakeholders and the way the model should work. 
	\item Data processing stage which involves data collection, cleaning and labelling (in case of supervised learning).
	\item Feature engineering stage which involves the modification of the selected data. 
	\item Model training stage which is related to the way the selected model is trained and tuned on the (labeled) data. 
	\item Model evaluation stage which regards to the measurements used in order to evaluate the model. 
	\item Model deployment stage which includes deploying, monitoring and maintaining the model. 
\end{itemize}


Nascimento et al. \cite{de2019understanding} conducted a survey with 6 Brazilian software companies reaching the conclusion about ML development following a 4-stage process in these companies (understanding the problems, data handling, model building, and model monitoring). 


The work of Banimustafa and Hardy \cite{IEEE9} is a practical application of a proposed scientific data mining process model in data mining, more specifically in metabolomics. The model was inspired by Software Engineering (among other fields) and although the paper proposes specific workflow stages (such as: data pre-processing, Data Exploration, Technique Selection, Knowledge Evaluation, Deployment, and Process Evaluation), the authors pointed out that the proposed framework could be generalized in order to perform data mining in other scientific disciplines.

The work of Gotz et al. \cite{ACM196} addresses the challenges that arise when trying to adopt traditional Software Engineering practices in Machine Learning Modeling, since the authors identified issues regarding the requirements design stage, as well as differences between traditional software systems and Machine Learning models’ lifecycles and workflows. 

The work of Hutchinson et al.  \cite{hutchinson2020towards} 
focuses on the Data Processing stage but also approaches other stages of the Machine Learning Workflow such as Requirements, Design, Implementation, Testing and Maintenance. The main goal of the author is to better understand the processes that generate the development of data and highlight the importance of adopting practices that enable accountability throughout the data development lifecycle.

Singla, Bose and Naik \cite{singla2018analysis} studied the logs related to software engineering following the agile methodology for a machine learning team and compared it with the logs for a non-machine learning team, analyzing the trends and their reasons. The authors then provided a few suggestions about the way Agile could have a better use for machine learning teams and projects. 



The work of Kriens and Verleben  \cite{kriens2019software} which is one of the few we found that was done from a Software Enginering perspective, also proposes a Machine Learning Workflow based/inspired by the stages of the Software Engineering lifecycle. However, the Worklflow’s stages of this work are different from the one proposed by Amershi et al. \cite{8804457}. 

The work by Lo et al.  \cite{lo2020systematic} is the part of the group consisting of a number of papers which have a software engineering perspective. Although the authors proposed a cyclical workflow for Federated ML (Background understanding, Requirements Analysis, Architecture Design, Implementation and Evaluation, and back to Background understanding), the article also mentions well-known machine learning stages such as: data collection, data pre-processing, feature engineering, model training, and model deployment. It is worth mentioning that the paper also deals with anomaly detection, which is a ML technique.
Among the findings, they have highlighted that the most discussed phase is model training. They have also found that “only a few studies cover data pre-processing, feature engineering, model evaluation, and only Google has discussions about model deployment (e.g., deployment strategies) and model inference. Model monitoring (e.g., dealing with performance degradation), and project management (e.g., model versioning) are not discussed in the existing studies.”

In the work of Hesenius et al. \cite{hesenius2019towards} the authors argue that developing ML/AI applications is typically a subproject of an overarching development cycle, thus feedback loops and connections are needed to integrate all activities. Consequently, they introduce their own proposed workflow for engineering data-driven applications, describe the roles team members take, and finally describe how different phases are structured, namely: Developing and Understanding the Application Domain;  Creating a Target Data Set; Data Cleaning and Pre-processing ,Data Reduction and Projection; Choosing the Data Mining Task; Choosing the Data Mining Algorithm; Data Mining; Interpreting Mined Patterns; and Consolidating Discovered Knowledge.

Rahman et al. \cite{rahman2019machine} presented an industrial case study, in which they apply machine learning (ML) to automatically detect transaction errors and propose corrections. The authors identified and discussed the challenges that they faced during this collaborative research and development project from three distinct perspectives: Software Engineering, Machine Learning, and industry-academia collaboration. In this way the work addresses the Software Engineering stages (requirements engineering, Design, Implementation, Integration, Testing, Deployment), Machine Learning Development Workflow stages (Problem Formulation, Data Acquisition, Pre-processing, Feature Extraction, Model Building, Evaluation, Integration and Deployment, Model Management, and AI Ethics) and Industry-academia Collaboration stages (Problem Understanding, Knowledge Transfer, Focus on objectives, Professional Practice, and Privacy and Security). It is also worth mentioning that the authors have adopted the Agile approach for Research and Development.

The work by Reimann and Kniesel-Wünsche \cite{reimann2020achieving} compare ML Workflows (with similar stages to the ones mentioned by Amershi et al.\cite{8804457} against traditional SE Workflows in order to address the lack of guidance in what currently used development environments and ML APIs offer developers of ML applications and contrast these with software engineering best practices to identify gaps in the current state of the art.

We have also found other articles focusing on specific stages of the Software Engineering lifecycle or the Machine Learning Workflow such as \cite{ACM173}, \cite{ACM43}, \cite{IEEE7}, and \cite{ACM179}. Some of these papers will be also addressed in the Gap section (RQ4).

\subsection{RQ2: What are the techniques applied in each of phases of the machine learning model development phases}
Although we did not find any mention of any specific techniques regarding model requirements stages and training stages, we were able to collect several insights about the stages of data processing, future engineering, model evaluation, and model deployment. 

According to Correia et al. \cite{Correia} the unique Data Processing method is the use of charts, such as box plots and histograms, to aid with the verification of data quality. Additionally, the main reason for the adoption of such visual tools is to avoid the use of inappropriate data so the data scientist can avoid the risk of increasing development costs by re-execution of Data Processing in case of error identification in later stages like Feature Engineering. 

Amershi et al. \cite{8804457} mention that Data Processing stage makes use of rigorous data versioning and sharing techniques , since Microsoft teams have found it necessary to blend data management tools with their ML frameworks to avoid the fragmentation of data and model management activities and also because the authors identified that a fundamental aspect of data management for machine learning is the fast pace in the evolution of data sources. In this way, continuous changes in data may arise either from (i) operations initiated by engineers themselves, or from (ii) incoming fresh data (e.g., sensor data, user interactions).  In an example about the application of such techniques provided by the authors, each model is tagged with a provenance tag that explains which data it has been trained on and which model version, and each dataset is tagged with information about where it originated from and which code version was used to extract it (and any related features). These techniques are used for mapping datasets to deployed models or for facilitating data sharing and reusability.

As for the Feature Engineering stage, Correia et al. \cite{Correia} mentioned that statistical methods in data analysis, and the use of automatic feature selectors in feature selection are the two main methods used for performing the Feature Engineering. 
Although the authors mention how statistical methods were widely used to assist the data analysis process to help data scientists observe data behavior, they did not mention any specific tools. 
Regarding the use of automated feature selector, it is important to demonstrate that there is a fine distinction between which use is associated with deep learning and which use is associated with other algorithms. In fact, the Feature Engineering Stage is skipped when dealing with deep learning algorithm (since algorithms for this purpose automatically learns the best feature for problem solving and model training, discarding the need for data scientists to do so), whereas when dealing with other kinds of algorithms, the feature selection is performed manually with data scientists executing operations like feature scoring to ranking features based on relevance.   
Regarding model Evaluation, Amershi et al. \cite{8804457} mention that Machine Learning centric software goes through frequent reviews initiated by model changes, parameter tuning, and data updates, and the combination of these has a significant impact on system performance. In this context, they have identified the use agile techniques to evaluate experiments and the use of combo-flighting techniques (such as flighting a combination of changes and updates), including multiple metrics in experiment score cards, and performing human-driven evaluation for more sensitive data categories in order to developed systematic processes.  

Finally, to ensure all aspects run seamlessly during Model Deployment the authors recommend the following:  (i) automating the training and deployment pipeline; (ii) integrating model building with the rest of the software; (iii) using common versioning repositories for both ML and non-ML codebases, and tightly coupling the ML and non-ML development sprints and standups.

\subsection{RQ3: What are the pros and cons of each machine learning model development technique?}

The issue covered in RQ3 (where there were a significant number of venues with a perspective different from our main research purpose, addressing the question) is more prominent here. In fact, we could only find limited answers in articles from a perspective of Machine Learning Techniques applied to Software Engineering tasks or applied to the stages of the Software Engineering Model Development life-cycle. In other words, we only found limited answers from articles that were mostly unrelated to our research purposes. 

In this way, 
the work of Hesenius et al. \cite{hesenius2019towards} discusses different Machine Learning Model Techniques from supervised to unsupervised  learning, but without specifying the pros and cons of each one.

Additionally, 
Nguyen et al \cite{nascimento2020software} search for the learning paradigms classification (mentioned in the development of ML systems of organization) by searching with keywords such as Supervised Learning, Unsupervised Learning, Reinforcement Learning among others.
In the work of Shafiq \cite{shafiq2020machine} the author showed interest in understanding whether a particular type/technique was consistently employed for a specific life cycle stage, bearing in mind that the ML technique refers to how the models have been trained, e.g., supervised, semi-supervised or unsupervised and how it is related to the algorithms such as a support vector machine (SVM), random forests (RF) or neural networks (NN). 

Articles reviewed by the author employed supervised learning, whereas 14 out of 227  articles employed unsupervised learning, and 6 out of 227  employed semi-supervised learning. Likewise, 4 out of 227  addressed reinforcement learning, 1 out of 227  focused on analytical (inference based) learning, while the rest of the articles (40 out of 227) reported none. 
Although all the ML techniques have certain pros and cons, the selection of the most suitable technique depends on the type of dataset being constructed or employed and the authors did not provide further information on that.

Finally, 
an interesting contribution we have found to this research question came from the article of 
Wang et al.\cite{wang2020synergy} summarizes the characteristics of each Machine Learning Model Development Technique by highlighting some of their respective advantages: supervised learning (SL), unsupervised learning (UL), semi-supervised learning (SSL) and reinforcement learning (RL). SSL presents a challenging learning setting, while in SL, the training data comprises examples (represented in the form of vectors). UL is often used to discover groups of similar examples and RL is concerned with the problem of finding suitable actions to take in a given situation in order to maximize a reward. According to the authors, by using an UL technique, the system’s performance may be unstable in comparison with supervised techniques.

\subsection{RQ4: What are the gaps in terms of the model development life-cycle?}

We have identified some gaps in terms of model development life cycle which were mentioned in the literature since the processes adopted by data scientists in their companies are non-linear, requiring too much rework to satisfy customers’ needs (Correia et al.\cite{Correia}). Additionally Correia et al. \cite{Correia} did not find any activities that address the verification and the validation of the artifacts generated during the workflow stages. As a consequence, the gaps identified in the developing stages were attributed to the existence particularities identified in the Machine Learning model development. According to the authors, practitioners should anticipate problems and save resources in order to mitigate recurrent feedback loops in the process.  Further, they state that the best way to accomplish that is by following software engineering practices starting from the early ML modeling stages, which not only allows companies to reduce rework and dependence on the domain experts, but also leverage the maintainability of ML models. For instance, they mentioned the importance of developing customized inspection techniques to support the verification of Machine Learning features and models.
This vision is somewhat supported by Amershi et al. \cite{8804457}, as when the authors state that data scientists perceive Data Processing and Feature Engineering as one of the more challenging stages of Machine Learning model development. They describe how ML development lacks the support of a well-engineered process, and that the validation of the ML model is often not done, given the difficulty to test back-box ML models.

In a similar fashion we have found other works addressing gaps in specific stages of Software Engineering Development Stages and Machine Learning Workflow. Among these articles we can highlight Meyer\cite{IEEE7}, Wan et al.\cite{IEEE17}, Wolf and Paine\cite{ACM179}, Foidl and Ferderer\cite{ACM173}, Gotz et al.\cite{ACM196}, Tsay et al.\cite{ACM43}, and Simmons et al.\cite{ACM30}.

We have identified that most of the stage-related gaps are connected to the difficulties in adopting Software Engineering lifecycle in Machine Learning Modeling. This view was corroborated for the works of Ishikawa and Yoshioka\cite{ACM17}, Khom et al. \cite{IEEE24}, and Kim\cite{IEEE14}.
By conducting a survey with 278 professionals with proven experience in ML or practical ML applications in Japan, the work of Ishikawa and Yoshioka \cite{ACM17} found that due to the unique nature of ML-based systems, they would need new approaches in terms of software development processes. Moreover, according to the authors, the attempts to address this were not enough to eliminate the gaps resulting from this difference.
According to the article of Khom et al. \cite{IEEE24}, the failures and shortcomings professionals and researcher have been experiencing with Machine Learning systems are due to the fact that the rules of software development do not applied in Machine Learning Modeling where the rules come from the training data (and from which the requirements are generated) representing an additional challenge in terms of model testing and model verification. 

Finally, the article written by Kim   \cite{IEEE14} highlights the difficulty of incorporating software engineering development processes into the Machine Learning workflow since data-centric software development, such as Machine Learning Models, would be significantly different from traditional software development, mostly regarding testing, debugging and the probabilistic characteristics of those systems.

\subsection{RQ5: What are the trends regarding the techniques applied in the machine learning model development life cycle?}

We have identified two distinct trends regarding Software Engineering Model Development life cycle applied to Machine Learning Model Development. 
The first trend states that the integration of SE development process into Machine Learning modeling must consider the intrinsic differences between Machine Learning based systems and other applications such as data dependency. This pattern is clear in the works presented in \cite{8804457,Correia,zhang2019empirical}.  

The second trend states that considering data dependency in machine learning model development leads to different processes/adaptations and partial solutions. Likewise, some articles defend the development of a single general Machine Learning framework (in contrast with specific solutions destined to solve particular and partial issues related to ML/AI modeling), regardless of the type of data under consideration. According to this trend, the particularities of Machine Learning Based Systems do not imply the need for different Software Engineering Processes.  We can consider the works in \cite{hesenius2019towards,serban2020adoption,washizaki2019studying} to be aligned with this view. 


\subsection{Discussion}\label{sec:discussion}
Most of the few existing approaches of SE to ML are focused on broadly ML workflows \cite{8804457,Correia,IEEE9,kriens2019software,lo2020systematic,hesenius2019towards,rahman2019machine,reimann2020achieving,de2019understanding,IEEE17,ACM179,ACM173,ACM196,ACM43,ACM30}.
However, many studies do not provide details for each stage of the workflow, nor describe the techniques and algorithms that were applied, nor provide an evaluation of their approach by discussing the pros and cons.
Because of this lack of details, there is also a need for specialized stage approaches, focusing on a specific step of the workflow. In particular, there is an evident lack of approaches to support the requirements and maintenance SE development stages. Table \ref{tab-summarize} provides a simplified view of our findings related to specific SE tasks.  

\begin{table*}[!htb]
\centering
\caption{Issues and Approaches reported in some of the Primary Studies.  }
\label{tab-summarize}
\scalebox{0.90}{
\begin{tabular}{|c|c|c|c|}
\hline
\rowcolor[HTML]{C0C0C0} 
SE process or practices                                              & Issues                                                                                                                                                                              & Solutions                                                                                                                                                                                                              & Primary Studies                                                               \\ \hline
Data  processing                                                     & \begin{tabular}[c]{@{}c@{}}- Lack of methods and tools to \\ support software engineers to \\ perform data validation\\ - Lack of tools to support data management\end{tabular} & \begin{tabular}[c]{@{}c@{}}- Decision support about data prioritization and rigor\\ - Adoption of visual tools\\ -Use of rigorous data  versioning and sharing techniques\\ -Provenance tag for data models\end{tabular} & \begin{tabular}[c]{@{}c@{}}\cite{ACM173,Correia,8804457}\end{tabular}          \\ \hline
Documentation and Versioning                                         & \begin{tabular}[c]{@{}c@{}}- Extract metadata\\  from repositories is difficult\end{tabular}                                                                                      & \begin{tabular}[c]{@{}c@{}}- Catalog of ML models to support design and \\ maintenance\end{tabular}                                                                                                                    & \begin{tabular}[c]{@{}c@{}}\cite{ACM43}\end{tabular}                       \\ \hline
Non-functional Requirements                                          & \begin{tabular}[c]{@{}c@{}}- security\\ -unassured reliability and lacking transparency\end{tabular}                                                                           & \begin{tabular}[c]{@{}c@{}}- Identify parts of the ISO 26262 to be adapted to ML\\ - An approach based on dependability assurances\end{tabular}                                                                        & \begin{tabular}[c]{@{}c@{}}\cite{ACM68,ACM215}\end{tabular}                       \\ \hline
\begin{tabular}[c]{@{}c@{}}Design and\\  Implementation\end{tabular} & \begin{tabular}[c]{@{}c@{}}- APIs look and feel like \\ conventional APIs, but \\ abstract away data-driven behavior\end{tabular}                                                 & \begin{tabular}[c]{@{}c@{}}- catalog of design patterns for ML development\\ - information to support documentation and design of \\ APIs\end{tabular}                                                                 & \begin{tabular}[c]{@{}c@{}}\cite{washizaki2019studying,ACM127}\end{tabular}       \\ \hline
Evaluation                                                           & \begin{tabular}[c]{@{}c@{}}- Testing interpretability, \\ privacy, or efficiency of ML\end{tabular}                                                                               & \begin{tabular}[c]{@{}c@{}}- Proposal of new test semantic\\ - Tests based on quality score\end{tabular}                                                                                                               & \begin{tabular}[c]{@{}c@{}}\cite{zhang2019empirical,ACM121,ACM62}\end{tabular} \\ \hline
Deployment and Maintenance                                           & \begin{tabular}[c]{@{}c@{}}- Lack of support to adapt based on feedback\end{tabular}                                                                                            & - An approach to support adaptation based on quality gates                                                                                                                                                             & \cite{ACM62}                                                                         \\ \hline
\begin{tabular}[c]{@{}c@{}}Software Capability Maturity\\ Model (CMM)\end{tabular}                     & \begin{tabular}[c]{@{}c@{}}-Convert business requirements into\\ data requirements\\- Support project management\\ (e.g. estimate data budget) \end{tabular}                                                                                            & - a maturity framework for ML based on CMM                                                                                                                                                             & \cite{akkiraju2020characterizing}  
 \\ \hline
\end{tabular}
}
\end{table*}


\section{Threats for Validity}

We identified three potential threats to the validity of our study and its results. First there may be bias from the articles which provided answers for most of the research questions. Although we have identified a relatively significant number of articles related to software engineering applied to machine learning modeling, only some of them are related to the machine learning workflow, very few of those are from the software engineering perspective and even fewer considering the application of Software Engineering Life Cycle to Machine Learning based systems. In order to mitigate this kind of bias we have worked with a larger number of articles that were not directly related to our subject but that could provide us with insights about our research questions.  

A significant part of the articles whose conclusions were significant for this research were based on survey answers which may be subject to bias. That is because some of them came from a single company or a small group of companies in the same geographical region. 

Finally, in regard with results validation, we have only identified one article which has applied robust quantitative measurement in order to determine (validate) the impact/importance of certain practices to the Machine Learning workflow. It is worth mentioning that these measurements were applied to evaluate the impact of practice adoption and not the methods adoption, or the impact from adopting SE life-cycle in Machine Learning modeling.


\section{Challenge for Future Research}

The results we have found lead us towards a new set of possibilities in terms of future work. First, we can reconduct our literature review by refining our search strategy in order to compare/check the results and look for new insights. Second, experiments/surveys like those conducted by Amershi et al. \cite{8804457} and Correia et al. \cite{Correia} could be replicated with more companies from different geographic locations. 

Maybe the greatest achievement would be to incorporate some robust quantitative measurement such as the regressions conducted in Serban et al. \cite{serban2020adoption} in order to determine the impact or degree of improvement achieved from the adoption of the Machine Learning workflow based on the stages of Software Development lifecycles. 

The study also could be extended in order to check the impact of Software Engineering Processes in systems  based on specific Machine Learning Algorithms (or in machine learning based systems characterized for using a specific type of data), not only to determine the real impact of adopting Software Engineering development lifecycle stages in machine learning modeling, but also decide if there is a need to adapt the machine workflow according to the machine learning algorithm (or type of data). Possibly this may answer the question of whether we need multiple workflows or can use a single more general workflow suitable for all kinds of machine learning models.

\section{Conclusions}


This systematic literature review promotes a better understanding about how Software Engineering Development lifecycle can improve/address the recurrent issues identified on Machine Learning Model Development. 
Among the expected contributions, it is important to point out that although we found some articles proposing Machine Learning Workflows with some differences from each other, we were able to draw/identify the most comprehensive which is aligned with the Software Engineering Model Development Stages. 

Another expected result was the highlighting of data dependency as the main characteristic of Machine Learning Models. This finding lead to the identification of the most challenging stages of Machine Learning Development Process as:  Data Processing and Feature Engineering. 
We have also identified two distinct trends regarding the results and the conducted research articles.  


Finally we concluded that the understanding of how Software Engineering Model Development practices and the adoption of a Machine Learning Workflow in accordance with those practices, more specifically, with the Software Engineering life-cycle is a subject of vital importance for the evolution of Machine Learning/Artificial Intelligence and continuing development of its applications (especially on a large scale), even if further research on the topic is needed.

\section*{Acknowledgment}
This work was supported by the Natural Sciences and Engineering Research Council of Canada (NSERC), and the Ontario Research Fund (ORF) of the Ontario Ministry of Research, Innovation, and Science, and the Centre for Community Mapping (COMAP).

\bibliographystyle{IEEEtran}
\bibliography{bib/bib-ieee,bib/bib-disciplina,bib/bib-acm,bib/bib-avulso}

\end{document}